\documentclass[prd,aps,twocolumn,showpacs,preprintnumbers,amsmath,amssymb,nofootinbib]{revtex4}
\usepackage[dvips]{graphicx}
\usepackage{epsf}
\usepackage{amsmath}
\usepackage{amssymb}

\usepackage{graphicx}
\usepackage{dcolumn}
\usepackage{bm}
\def\be{\begin{equation}}
\def\ee{\end{equation}}
\def\bea{\begin{eqnarray}}
\def\eea{\end{eqnarray}}

\begin{document}

\title{Primordial perturbation with a modified dispersion relation}

\author{Yi-Fu Cai$^{1}$ and Xinmin Zhang$^{1,2}$}

\affiliation{1) Institute of High Energy Physics, Chinese Academy
of Sciences, P.O. Box 918-4, Beijing 100049, P.R. China}



\affiliation{2) Theoretical Physics Center for Science Facilities
(TPCSF), Chinese Academy of Sciences, P.R. China}


\begin{abstract}
In this paper we study the generation of primordial perturbation
with a modified dispersion relation in various cosmological
evolutions. We stress that the formation of the power spectrum is
strongly dependent on the background. Working in a bounce model
with a matter-like contracting phase, we obtain a red tilt
spectrum due to the modified dispersion relation.
\end{abstract}

\maketitle

\section{Introduction}

The nature of primordial perturbation\cite{Bardeen:1980kt},
responsible for the formation of large scale structure of the
universe, has been intensively studied in the literature. The
cosmological observations favors an adiabatic and nearly
scale-invariant spectrum of primordial
perturbation\cite{Bardeen:1983qw}, as predicted by an inflation
model \cite{Guth:1980zm} with a standard dispersion relation.
However, advocated by the development of quantum gravity, one may
question on the initial state of inflation and the perturbations
generated in this model. As a possible consequence of quantum
gravity, the dispersion relation of primordial perturbation may be
modified\cite{ArmendarizPicon:2003ht, ArmendarizPicon:2006if,
Piao:2006ja, Magueijo:2008sx}.

The studies around modified dispersion relations were earlier
addressed in DBI inflation\cite{Silverstein:2003hf} by introducing
a sound speed parameter, which has shown plentiful
phenomenons\cite{Chen:2005ad, Cline:2006hu, Cai:2008if,
Cai:2009hw}. A modified dispersion relation can also be derived in
the models of modified gravity, such as trans-Planckian
physics\cite{Martin:2000xs}, noncommutative field
approach\cite{Cai:2007xr}, loop quantum
gravity\cite{Bojowald:2006zb}, Lorentz-violating
models\cite{Jacobson:2000gw, ArkaniHamed:2003uy, Cai:2007gs}, and
so on. Recently, within the frame of a non-relativistic gravity
model\cite{Horava:2009uw}, several works have appeared to claim
that a scale-invariant spectrum can be obtained without carefully
considering the background cosmological
evolution\cite{Mukohyama:2009gg} and even without matter
components\cite{Cai:2009dx, Chen:2009jr}; however, also see Refs.
\cite{Piao:2009ax, Kim:2009zn, Gao:2009ht} for reexaminations.
This is an interesting issue in which a modified dispersion
relation was applied, but one ought to take the background
cosmological evolution into account seriously. Assuming a mode of
cosmological fluctuations emerges at the beginning of the
universe, the form of its initial condition strongly depends on
the background. Moreover, if this mode is responsible for the
formation of the large scale structure at late times, it has to
experience a process of exiting the Hubble scale in aim of
freezing its oscillating behavior\cite{Mukhanov:1990me}. This
process is also determined by the background evolution.

In the current paper, we study the generation of the primordial
perturbation by introducing a modified dispersion relation
phenomenologically. Through following one mode of these
perturbations along with the cosmological evolutions, we show
that, a modified dispersion relation for primordial perturbations
can not make their generations be independent on the background
evolution, but changes the way of their freezing. Especially, if
the universe is preceded by a bounce\cite{Cai:2007qw}, the method
of seeding a scale-invariant spectrum with a modified dispersion
relation can only be applicable in a fine-tuned regime.

This paper is organized as follows. In Section II we show how the
primordial fluctuations emerged inside the Hubble radius can seed
the cosmological inhomogeneities when the dispersion relation is
modified. In Section III we perform the detailed calculation of
primordial perturbations by introducing a modified dispersion
relation phenomenologically. Specifically, we study the spectrum
in a pure expanding universe and a bouncing universe respectively.
Our results show that the spectrum obtained in a pure expanding
phase can be spoiled by a cosmological
bounce\cite{Brandenberger:2009rs}. In Section IV we study the
primordial perturbation with a modified dispersion relation in a
specific bounce model and obtain a red tilt power spectrum.
Section V is our conclusion and discussions.

\section{Seeding primordial perturbations}\label{sec:spp}

We begin with a brief discussion on the cosmological evolution of
primordial perturbations in the frame of an expanding universe.
Due to the fact of cosmological inhomogeneities observed in the
universe, it requires that the cosmological fluctuations initially
emerge inside a Hubble radius, and then leave it in the primordial
epoch, and finally reenter at late times. In an expanding
universe, with a standard dispersion relation, this process can
usually be realized by stretching the physical wavelength
$\frac{a}{k}$ longer than the Hubble radius $1/H$. One may take
the scale factor as
\begin{eqnarray}\label{scalefactor}
a(t) = a_0 \bigg(\frac{t}{t_0}\bigg)^p~,~ {\rm with}~~
p=\frac{2}{3(1+w)}~,
\end{eqnarray}
where $w$ is the background equation-of-state, and the subscript
``$0$" denotes any fixing point along the cosmological evolution.
Then the Hubble parameter takes the form
\begin{eqnarray}
H=\frac{p}{t}~.
\end{eqnarray}

From the above equations, one can find that, for a mode of
perturbation with the standard dispersion relation to exit the
Hubble radius in an expanding universe, there has to be the
background equation-of-state $w<-\frac{1}{3}$, such as inflation.
It may be characterized by an efolding number parameter, which is
defined as,
\begin{eqnarray}
{\cal N}\equiv\ln\bigg(\frac{k_f}{k_i}\bigg)~,
\end{eqnarray}
where $k$ is the comoving wave number, and the subscript ``$i$"
and ``$f$" denotes the earliest and last modes generated in
primordial epoch. For example, from current observations, this
parameter is required to be around $60$.

However, the dispersion relation may be modified at high energy
scales, for example, due to trans-Planckian physics, Lorentz
violating effects. If it happens, the scenario of the above
generation for the primordial perturbation has to be changed
correspondingly. Now we phenomenologically introduce a modified
dispersion relation
\begin{eqnarray}\label{mdf}
\nu=kf(k_{ph})~,
\end{eqnarray}
with
\begin{eqnarray}
f(k_{ph}) =
  \left\{ \begin{array}{c}
          (\frac{k_{ph}}{M})^\alpha,~~k_{ph}>M \\
          \\
          1,~~k_{ph} \leq M
\end{array} \right.
\end{eqnarray}
where $k_{ph}=\frac{k}{a}$ is a physical wave number and assume
$\alpha>0$. We suggest there is an energy scale $M$ for the
modification, so that the standard perturbation theory can be
recovered at low energy scales.

Suppose we still work in the frame of an expanding universe. In
order to realize the modes with a modified dispersion relation
(\ref{mdf}) to exit the Hubble radius, it requires that
$M^{\alpha}(\frac{a}{k})^{1+\alpha}$ grows faster than the Hubble
radius $\frac{1}{H}$. Therefore, there is a condition
\begin{eqnarray}\label{condition1}
w<-\frac{1}{3}+\frac{2\alpha}{3}~,
\end{eqnarray}
for the background equation-of-state. One may notice that, the
modes are much easier to escape the Hubble radius when $\alpha>0$,
and thus it could be unnecessary to require an inflationary period
at early times. For a recent gravity model, proposed by
Ho\v{r}ava, $\alpha$ is equal to $2$ and the perturbations are
able to exit the Hubble radius just requiring $w<1$. In this case,
it seems that requirements on the background evolution becomes
quite loose, but this claim is not suitable. At least, we need the
universe is expanding with all the perturbation modes emerge at a
finite initial moment.

\begin{figure}[htbp]
\includegraphics[scale=0.5]{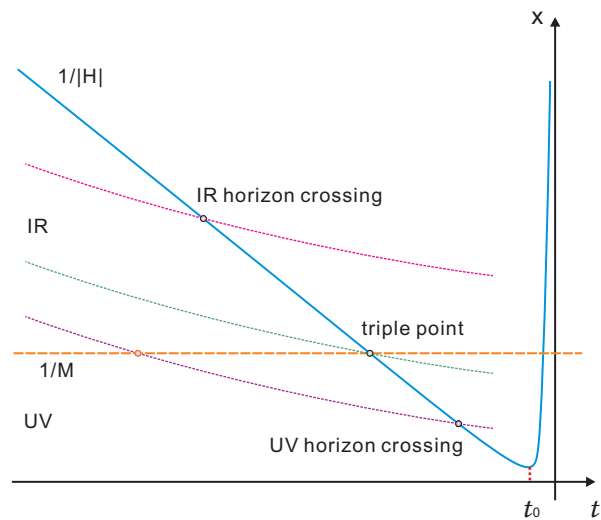}
\caption{A sketch plot of the cosmological evolution for the
primordial perturbations with a modified dispersion relation.}
\label{fig:sketch}
\end{figure}

Moreover, if the initial big bang singularity is replaced by a big
bounce and the universe has experienced a contracting phase, the
evolution of primordial perturbations will be changed
totally\footnote{A cosmological bounce may spoil the usual results
obtained in an inflation model as studied in Refs.
\cite{Cai:2007zv, Cai:2008qb, Cai:2008ed}.}. In this case, the
scale factor is very large initially and so the modes of
perturbation are almost in the infrared (IR) regime with
$k_{ph}\leq M$. Along with the contraction, these modes may
firstly enter the ultraviolet (UV) regime with $k_{ph}>M$, and
then exit the Hubble radius; or, they exit the Hubble radius from
the IR regime directly. We give a sketch plot to describe this
scenario in Figure \ref{fig:sketch}. From the figure, one can read
that there are three critical moments,
\begin{eqnarray}
 t_{IR}(k) &=& t_0^{\frac{p}{p-1}} \bigg( \frac{k}{pa_0} \bigg)^{\frac{1}{p-1}}~, \nonumber\\
 t_{UV}(k) &=& t_0^{\frac{p(1+\alpha)}{p(1+\alpha)-1}}
  \bigg( \frac{k^{1+\alpha}}{pM^{\alpha}a_0^{1+\alpha}} \bigg)^{\frac{1}{p(1+\alpha)-1}}~, \nonumber\\
 t_{UI}(k) &=& t_0 \bigg(\frac{k}{Ma_0}\bigg)^{\frac{1}{p}}~,
\end{eqnarray}
which are IR horizon crossing time $t_{IR}$, UV horizon crossing
time $t_{UV}$, and UV/IR crossing time $t_{UI}$ respectively.
These three moments overlap at a triple point
$t_{tp}=\frac{p}{M}$, and correspondingly the comoving wave number
of the critical mode takes the value
$k_{tp}=a_0M^{1-p}p^pt_0^{-p}$. Only for the modes with
$k>k_{tp}$, the initial conditions could be changed by the
modified dispersion relation; while for those with $k\leq k_{tp}$,
their primordial spectrum is the same as that obtained with a
standard dispersion relation\cite{Brandenberger:2009yt}.

The condition for the generation of perturbation in a contracting
universe is exactly opposite to that in an expanding universe as
shown in Eq. (\ref{condition1}). It requires
$w>-\frac{1}{3}+\frac{2\alpha}{3}$ for the perturbation escaping
the Hubble radius in a contracting phase. Therefore, the case of
$\alpha=2$ suggests an Ekpyrotic phase is needed for the
generation of primordial spectrum. Besides, in order to make the
modes with $k > k_{tp}$ relevant for current observable ones, the
evolution of the universe can not be symmetric to the bounce
point. Namely, one may introduce an inflationary stage after the
bounce to stretch the UV modes into IR regime on super-Hubble
scales.

\section{The spectrum of primordial perturbation with a modified dispersion relation}

To make the above analysis specific, we do the calculation of the
spectrum of primordial perturbation with a modified dispersion
relation. To seed the fluctuations, we consider a test scalar
field $\phi$ which does not contribute to the background
evolution, but only provides a degree of freedom for scalar
perturbations $\delta\phi$. In the Fourier space, the equation of
motion for its fluctuation $\delta\phi$ is given by
\begin{eqnarray}\label{eom}
v_k''+(\nu^2-\frac{a''}{a})v_k=0~,
\end{eqnarray}
where the variable $v_k$ is defined by $v_k\equiv a\delta\phi_k$,
and the prime denotes the derivative with respect to the comoving
time $\eta\equiv\int\frac{dt}{a}$. For a constant
equation-of-state $w$, there is
\begin{eqnarray}\label{mass}
\frac{a''}{a}=\frac{\gamma}{\eta^2}~,~~{\rm
with}~~\gamma=\frac{p(2p-1)}{(1-p)^2}~.
\end{eqnarray}

Eq. (\ref{eom}) has an asymptotic solution when we neglect the
last term $\frac{a''}{a}$ which implies $|\nu\eta|\gg1$, and it is
strongly oscillating like trigonometric functions. This feature
coincides with an adiabatic condition $|\nu'/\nu^2|\gg1$, which
corresponds to the case that the effective physical wavelength is
deep inside the Hubble radius. Therefore, the modes can be
regarded as adiabatic when they are staying in the sub-Hubble
regime with $|\nu\eta|\gg1$, and we may impose a suitable initial
condition in virtue of WKB approximation,
\begin{eqnarray}\label{inicond}
v_k^{i}\simeq\frac{1}{\sqrt{2\nu(k,\eta)}}e^{-i\int^\eta\nu(k,\tilde\eta)d\tilde\eta}~,
\end{eqnarray}
for cosmological fluctuations.

We would like to analyze the effective mass term $\frac{a''}{a}$
closely. As discussed in Sec. \ref{sec:spp}, the generation of
primordial fluctuations strongly depends on the background
evolution. In suitable environments, the variable $|\nu\eta|$
could decrease along with the expansion. Once there is
$|\nu\eta|\ll1$, the modes will exit the Hubble radius, and the
equation of motion yields another asymptotic solution of which the
leading term takes
\begin{eqnarray}\label{sollead}
v_k^{l} \sim \eta^{\frac{1}{2}} \bigg[
c(k)\eta^{-\frac{1}{2}|\frac{1-3p}{1-p(1+\alpha)}|} \bigg]~.
\end{eqnarray}

From the above expression, one may observe that this solution can
seed appropriate perturbation spectrum at super-Hubble regime. Now
we match two asymptotic solutions (\ref{inicond}) and
(\ref{sollead}) at the moment of Hubble radius crossing
$|\nu\eta|\sim1$, and thus determine the form of $v_k$ at
super-Hubble scales,
\begin{eqnarray}\label{solution}
v_k(\eta) \simeq \frac{1}{\sqrt{2\nu}}
\bigg(\nu(k,\eta)\eta\bigg)^{\frac{1}{2}-\frac{1}{2}|\frac{1-3p}{1-p(1+\alpha)}|}~.
\end{eqnarray}
Applying the solution (\ref{solution}) directly, we obtain the
primordial spectrum as follows,
\begin{eqnarray}\label{spectrum}
P_{\delta\phi}&=&\frac{k^3}{2\pi^2}|\frac{v_k(\eta)}{a}|^2~\nonumber\\
  &\sim& \left\{ \begin{array}{c}
               k^{\frac{2-\alpha}{1-p-\alpha p}}~,~~(1-3p)[1-p(1+\alpha)]\geq0 \\
               \\
               k^{\frac{4+\alpha-6p(1+\alpha)}{1-p-\alpha
               p}}~,~(1-3p)[1-p(1+\alpha)]<0~.
\end{array} \right.
\end{eqnarray}

Finally, we obtain the power spectrum of primordial perturbations
without considering any specific models. From the result
(\ref{spectrum}), we find out that there are four sufficient
conditions for the spectrum to be scale-invariant. They are:
\begin{itemize}
\item (a) $p\rightarrow+\infty$ in the first case, which
corresponds to an inflation model\cite{Guth:1980zm};
\item (b) $p\rightarrow-\infty$ in the first case, which may be
realized in island cosmology\cite{Dutta:2005gt, Piao:2005na};
\item (c) $\alpha=2$ in the first case, and this case coincides
with the result obtained in the model of Ho\v{r}va-Lifshitz
cosmology\cite{Calcagni:2009ar, Kiritsis:2009sh};
\item (d) $p=\frac{4+\alpha}{6(1+\alpha)}$ in the second case.
\end{itemize}
In the following we will study the spectrum of primordial
perturbation in concrete background evolutions. Specifically, we
consider a purely expanding universe and a bouncing universe
respectively.

\subsection{In an expanding universe}

For a pure expanding universe, a mode of cosmological perturbation
emerge inside one Hubble radius when the universe was born, and
then it is able to enter the super-Hubble regime when the
condition (\ref{condition1}) is satisfied. This condition implies
$p>\frac{1}{1+\alpha}$ or $p<0$. Obviously, the first three
conditions can be realized in an expanding universe naturally. The
last condition can also be satisfied when we choose $\alpha>2$ and
$\frac{1}{1+\alpha}<p\leq\frac{1}{3}$.

\subsection{In a bouncing universe}

Now we work in the frame of bouncing cosmology which may be
realized by a model of non-relativistic gravity, namely in
\cite{Cai:2009in}. In this scenario the scale factor $a$ could be
quite large at early times, and so all the modes of primordial
perturbation stay in the IR regime initially, as shown in Figure
\ref{fig:sketch}.

As is analyzed in Section \ref{sec:spp}, a mode with comoving wave
number $k\leq k_{tp}$ can escape outside the Hubble radius in the
IR regime directly if $w>-\frac{1}{3}$. In this case, the
dispersion relation is in standard form $\nu=k$ and so the
spectrum can be scale-invariant only in a model of matter bounce
with $p=\frac{2}{3}$ \cite{Wands:1998yp, Finelli:2001sr} (see also
\cite{Starobinsky:1979ty}).

However, for the modes with $k>k_{tp}$, the condition of
generating primordial perturbation is
$w>-\frac{1}{3}+\frac{2\alpha}{3}$ and so requires
$0<p<\frac{1}{1+\alpha}$. Therefore, the sufficient condition (c)
implies that a scale-invariant spectrum in UV regime can be
obtained when $\alpha=2$ and $w>1$. Moreover, we can take
$\alpha<2$ and $\frac{1}{3}<p<\frac{1}{1+\alpha}$ to give a
scale-invariant spectrum in virtue of the condition (d).

The conditions leading to scale-invariant power spectra in an
expanding or contracting phase are summarized in Table I as shown
in the following.

\begingroup
\begin{table*}
\begin{tabular}
{||p{3.2cm}|p{2.5cm}|p{2.8cm}|p{2.5cm}|p{2.5cm}||}
\hline Background       &  Expanding                                                 & Contracting                       \\
\hline $\alpha=0$   IR  & $p\rightarrow\pm\infty$                                    & $p=\frac{2}{3}$                   \\
\hline $0<\alpha<2$ UV  & $p\rightarrow\pm\infty$                                    & $p=\frac{4+\alpha}{6(1+\alpha)}$  \\
\hline $\alpha=2$   UV  & $p>\frac{1}{1+\alpha}$ or $p<0$                            & $0<p<\frac{1}{1+\alpha}$          \\
\hline $\alpha>2$   UV  & $p=\frac{4+\alpha}{6(1+\alpha)}$ or $p\rightarrow+\infty$  &   N/A                             \\
\hline
\end{tabular}
\caption{Various possibilities of generating a scale-invariant
primordial spectrum in an expanding or contracting universe.}
\end{table*}
\endgroup

\section{A featured power spectrum in a bounce model}

In this section we will study the cosmological perturbation with a
modified dispersion relation in a bounce
model\cite{Brandenberger:2009yt}. Specifically, we consider the
model with an universe evolving from a matter-like contracting
phase\cite{Cai:2008qw}, of which the equation-of-state takes $w=0$
and so gives $p=\frac{2}{3}$. In this model, the perturbation with
a standard dispersion relation has been studied in details as
shown in Refs. \cite{Cai:2008ed, Cai:2008qw} and its
non-Gaussianities were studied by Refs. \cite{Cai:2009fn,
Cai:2009rd}. The modes in IR regime coincides with them, of which
the spectrum is as follows\cite{Cai:2009fn},
\begin{eqnarray}
P_{\delta\phi}^{\rm IR}=(\frac{H_0}{4\pi})^2~,
\end{eqnarray}
where we have placed the subscript ``$0$" around the bounce point
as shown in Figure \ref{fig:sketch}.

\begin{figure}[htbp]
\includegraphics[scale=0.8]{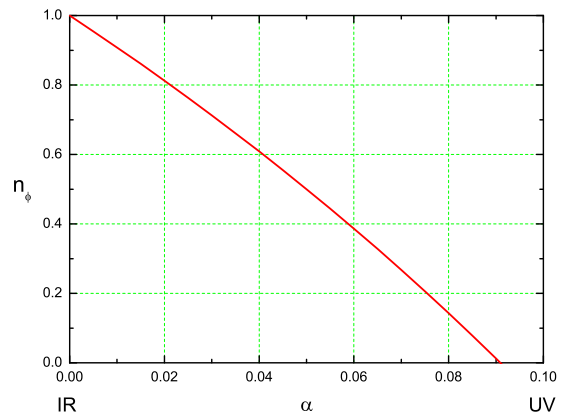}
\caption{The spectral index of the primordial perturbations with a
modified dispersion relation as a function of the parameter
$\alpha$.} \label{fig:ns}
\end{figure}

Now we deal with the UV modes of which the comoving wave numbers
satisfy $k>k_{tp}$. In order for the UV modes leaving the Hubble
radius, the parameter $\alpha$ has to be less than $\frac{1}{2}$.
According to Table I, we can find that these modes can not give
rise to an exactly scale-invariant spectrum. Consequently, we turn
to resolve the basic equation of motion for the perturbation
directly, and obtain the following solution
\begin{eqnarray}
v_k^{\rm
UV}\simeq\frac{1}{\sqrt{2\nu}}(\nu\eta)^{-\frac{1+\alpha}{1-2\alpha}}~,
\end{eqnarray}
in UV regime. From this solution, one can read that the modes at
super-Hubble scales are growing. This growth would stop when the
bounce takes place, and thus we can calculate the power spectrum
around the bounce point, which is given by,
\begin{eqnarray}
P_{\delta\phi}^{\rm UV} =
\frac{1}{2^{\frac{2(2-\alpha)}{1-2\alpha}}\pi^2}
H^{\frac{2(1+\alpha)}{1-2\alpha}} M^{\frac{3\alpha}{1-2\alpha}}
\bigg(\frac{a_0}{k}\bigg)^{\frac{9\alpha}{1-2\alpha}}~.
\end{eqnarray}
The corresponding spectral index can be derived as follows,
\begin{eqnarray}
n_\phi\equiv1+\frac{d\ln P_{\delta\phi}}{d\ln
k}=1-\frac{9\alpha}{1-2\alpha}~,
\end{eqnarray}
which is red tilt in UV regime. In the following, we plot the
spectral index of the primordial perturbation in this model in
Figure \ref{fig:ns}. One may notice that, in order to fit the
cosmological observations, for example Wilkinson Microwave
Anisotropy Probe (WMAP) data\cite{Komatsu:2008hk}, the parameter
$\alpha$ can not deviate from zero too much which provides a
strong constrain on models of modified gravity. And these results
indicate that a slightly red spectrum can be obtained by making a
very small modification on the dispersion relation in a matter
bounce model.

Another interesting signal concerns that, there could be a break
at the critical point of the power spectrum which is
phenomenologically similar to the one found by Ref.
\cite{Li:2009cu}.

\section{Conclusions and Discussion}

Modifications on the dispersion relation of a cosmological scalar
field have explored new approaches to generating the primordial
perturbations in various background evolutions. In the current
paper we have studied in details on the conditions for these
perturbations to form scale-invariant spectra. Especially, we
point out that the generation of primordial spectrum can not be
independent on the background evolution even in the model of
Ho\v{r}ava gravity with $\alpha=2$ in our case. These
perturbations may be transferred to the curvature perturbation at
late times by certain mechanisms, such as
curvaton\cite{Lyth:2001nq, Li:2008fma} or modulated
reheating\cite{Dvali:2003em, Kofman:2003nx}, and so responsible
for the structure formation of the universe. By investigating the
perturbation with a modified dispersion relation in a matter
bounce model, we show that the modification on the dispersion
relation gives rise to a red tilt spectrum as favored by current
observations.

Finally, we would like to comment on the choice of the parameter
$\alpha$ in the modified dispersion relation. In this paper, all
the results are based on an assumption that $\alpha$ is positive.
When relaxing this assumption we would expect to obtain more
interesting phenomenons, which deserves a detailed study in the
future.

\begin{acknowledgments}
We would like to thank Robert Brandenberger, Yun-Song Piao, Wei
Xue, and Yi Wang for discussions. The work is supported in part by
the National Science Foundation of China under Grants No. 10533010
and 10675136, and by the Chinese Academy of Sciences under Grant
No. KJCX3-SYW-N2.
\end{acknowledgments}

\end{document}